\documentclass{IEEEcsmag}

\usepackage[colorlinks,urlcolor=blue,linkcolor=blue,citecolor=blue]{hyperref}

\usepackage{upmath}
\usepackage{longtable}
\usepackage[table,xcdraw]{xcolor}

\jvol{XX}
\jnum{XX}
\paper{8}
\jmonth{May/June}
\jname{IEEE Software}
\pubyear{2020}

\usepackage{cleveref}

\begin{document}


\title{Mind the Gap: On the Relationship Between Automatically
Measured and Self-Reported
Productivity}

\author{Moritz Beller$^*$}
\affil{Facebook, Menlo Park, USA}

\author{Vince Orgovan, Spencer Buja}
\affil{Microsoft, Redmond, USA}

\author{Thomas Zimmermann}
\affil{Microsoft Research, Redmond, USA}
\markboth{}{Mind the Gap: On the Relationship Between Automatically Measured and Self-Reported Productivity}

\begin{abstract}
To improve software developers' productivity has been the holy grail of
software engineering research. But before we can claim to have improved
it, we must first be able to measure productivity. This is far from
trivial. In fact, two separate research lines on software engineers'
productivity have co-existed almost in complete isolation for a long
time: automated product and process measures on the one hand and
self-reported or perceived productivity on the other hand. In this
article, we bridge the gap between the two with an empirical study of 81
software developers at Microsoft.
\end{abstract}

\maketitle

\stepcounter{footnote}\footnotetext{This work was performed while Moritz was an intern at Microsoft Research,
Redmond, USA, and a researcher at Delft University of Technology, The Netherlands.}

\chapterinitial{The textbook} definition of productivity is the ratio of the output
product over the input effort that made it~\cite{ieee}. While this definition
sounds simple, its application to knowledge workers such as software
engineers has turned out to be a major challenge~\cite{wagner}---so much so,
that even after more than 40 years of work on productivity for
software engineers, we still have no generally agreed-upon way to
define, let alone measure productivity.

Two main methods on assessing productivity have thus far emerged: 1)
The automated measurement of product or process features to assess
productivity, such as the lines of code written per
day~\cite{petersen}, sometimes called traditional or ``objective''
productivity measures \cite{murphy}, and 2) the reporting of
self-assessed productivity by the software engineers \cite{murphy},
sometimes called self-assessed, self-rated, or self-perceived
productivity. Both metrics have distinct benefits and
shortcomings. While other metrics such as peer assessment exist, they
are much less widely studied.

An advantage of product or process metrics is that they are easy to
measure and give seemingly objective results. We refrain from calling
them objective, however, because: 1) The choice of which measures to
use is itself subjective. 2) The interpretation of a metric such as
lines of code per day is problematic and subjective---what about
engineers who produce more concise code in the same time, what about
that one-liner patch that took a week to debug and fixes a
business-critical problem? 3) Automatic measures can be gamed
\cite{murphy}.  4) In many cases, automatic measures fail to capture
work that happens off-screen, for example, meetings and
discussions. Addressing some of these issues, IBM introduced the
concept of Function Points in the late 1970s~\cite{fps}. However,
these, too, have drawbacks: they include an element of subjectivity by
how one judges ``functionality'' into the measurement itself.

In the other line of thought on productivity, software developers
therefore assess their own productivity instead~\cite{meyer}, called
self-reported productivity. Self-reported productivity circumvents some
of the challenges of automated measures, but it is susceptible to, for
example, cognitive biases \cite{murphy}. In contrast to many automatic
measures, its subjectivity also makes it hard to compare across
individuals.

We explicitly do not want to claim that one of these two ways to measure
productivity is inferior to the other. Both are widely-used measures
with their respective advantages and disadvantages. Until now, however,
they have existed in almost complete isolation of each other. Bridging
the gap between self-reported and automatically measured productivity
allows software engineers, managers, and researchers better understand
what makes them (feel) productive at work. In this article, we provide
the initial steps towards bridging this gap. In particular, we answer
the question how self-reported productivity relates to measured
productivity.

\section{Data Collection}\label{study-setup}

\noindent
To gain insight into what drives productivity and its perception, we
asked a sample of 1,066 Microsoft Windows developers to participate in
a study in which they rated their own productivity daily. We filtered
employees to reach out to by their job title being ``Software Engineer
I,'' ``Software Engineer II,'' or ``Senior Software Engineer.'' These
are the roles where the bulk of programming happens at Microsoft. Of
the 1,066 developers, 81 (7.6\%) finally participated in our
study. Participants reported an average mean of 8 years of
professional software development experience (median 6, minimum 1,
maximum 26 years), mostly within Microsoft. We wanted make the
participant pool as large as possible, so we followed an opportunistic
sampling strategy, allowing every developer who was interested to
participate. To further make participation more attractive, we
promised a personalized productivity report at the end of the study
and raffled three 50\$ Amazon vouchers.  During the 5 week study
period, developers submitted 1,479 daily productivity ratings in total
(out of 2,050 theoretically possible, if everyone submitted every
day). However, participants were not required to submit ratings daily,
for example because they took a day off.

We enrich these daily ratings with several product and process metrics
computed from hundreds of terabytes of data streamed from Microsoft's
internal data store system COSMOS. This included data from the Version
Control Systems as well as Windows and IDE usage telemetry. The nature
of the IDE telemetry data was comparable, but more abstract than known
from other systems, for example WatchDog~\cite{beller}.

\begin{figure*}
  \centering
\includegraphics[width=\textwidth]{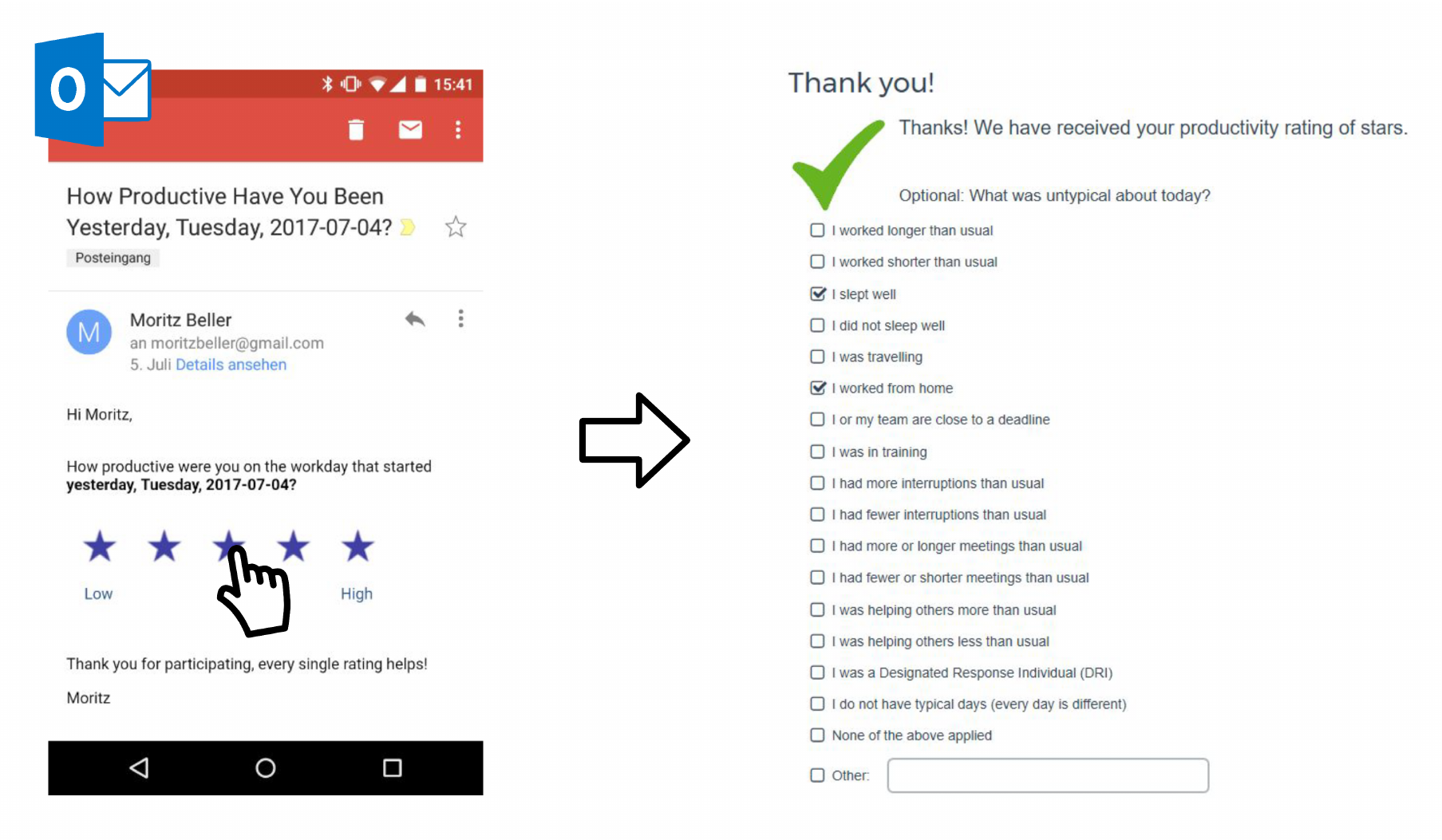}
\caption{Once a developer rated their productivity in the daily email
probe (left), they could enter additional attributes for the day, such
as ``I slept well'' and ``I worked from home'' (right).}
\label{mail}
\end{figure*}

\subsubsection{Daily Email Probes.}
At the heart of the study was a daily intervention with an email (a)
asking study participants to self-assess their productivity for the
day, depicted in \textbf{\Cref{mail}}.  Participants could choose when
they wanted to receive these emails, with the intention that they
could reflect on productivity at the end of their workday. We
maximized usability of the emails for multiple email clients. In
particular, we focused on mobile devices so developers could rate
their day for example during their commute. \Cref{mail} (left) shows
an example email of how developers rated their productivity on a five
star scale. We picked this scale for familiarity with popular other
rating systems such as Amazon's or Airbnb's. We also wanted to keep
the cognitive load small so that they would not need to differentiate
between a 7 or an 8 star rating for a day. Clicking on a star would
register the rating and bring participants to a web page (\Cref{mail},
right), on which they could further specify 17 attributes for a day or
enter an attribute of their own. Separate research found that these
pre-defined attributes distinguish a good from a bad day for a
developer~\cite{meyer3}, such as \emph{``I did not sleep well,'' ``I
  had fewer interruptions than usual,'' or ``I had more meetings than
  usual.''}

\subsubsection{Windows Telemetry.} With participants' consent, we
collected application usage data from telemetry built into internal
development builds of Microsoft Windows. This resulted in detailed
data on how much time developers spent in different applications on
their machines. Many developers at Microsoft use a desktop workstation
as well as a notebook; our telemetry gathers information from both and
identifies that they come from the same employee. This
state-of-the-art internal telemetry framework includes features such
as automatic detection of inactivity~\cite{beller}.  Similar to
previous work at Microsoft~\cite{meyer2}, we categorized applications
into a set of different classes, aiming to have at least the 90th
percentile of observed applications grouped into classes. This means
that in less than 5\% of applications, we would report time in a
non-standard application spent as ``Other'' when it really should be in
one of the pre-defined categories. For example, we classified
Microsoft Outlook and Mozilla Thunderbird into the email class and
Visual Studio, Eclipse, vim, emacs, and many others into the coding
class, but an exotic code editor only used by one developer might be
miss-labeled as other. This also bears the risk of misclassification if
a developer uses emacs to do their emails, since we would always label
it as ``coding''. However, we anecdotally know that such cases are
rare in the tightly-integrated Microsoft ecosystem. Moreover, the
classification was done by a member of the Windows developers sub-group
that we studied, so they are familiar with the typical tools and
processes. We then aggregated the amount of time spent in each
application class for every day. This way, we ended up with \emph{time
  spent on pure coding, debugging, code reviewing, testing, and other
  development tasks}.  Moreover, we also incorporated the \emph{time
  spent on emails, in the browser, and other applications}. The
\emph{total active time} summarizes the overall time spent active in
front of the computer for all applications. Monitoring developers'
application usage bears potential privacy risks. Therefore, respecting
developers' privacy was paramount in this study. We designed and
adhered to an ethics board approved study protocol with early
anonymization and did not store developers' personalized reports.

\subsubsection{Other data sources.} Other data sources included the code
review and version control systems at Microsoft. From them, we extracted
how many check-ins developers did in merged pull requests and how many
code reviews they performed per day. Moreover, we extracted all meetings
and events from their public Outlook calendars. We aggregated this
information only for the days developers took part in our study,
summarizing the duration of all events in the calendar the developer
accepted as the number of meetings in a day, the cumulative meeting
time, and whether the day was part of a meeting spanning multiple days.

\section{Modeling Self-Reported Productivity}\label{regression-models}

\noindent
The main goal of this study is to understand whether, and if so, which
automatically measurable factors could describe self-reported
productivity. To this aim, once we had collected all data, we started
a step-wise process of building and refining linear models with the
self-reported productivity score as a dependent variable and
combinations of telemetry and other data as the independent
variables. We chose linear models because they are the most basic form
of modeling and the easiest to understand. To get an initial overview of
relationships, we checked our data for the presence of
cross-correlations. In the following, we describe the results of the
different models we built in terms of the variance of the productivity
score we could explain with them (the so-called $R^2$
value). Explained variance is the standard metric to evaluate how well
a model fits the data---a perfectly fit linear model would, given the
input of our independent variables such as telemetry, always predict the
correct productivity score for any given day and developer.

\begin{enumerate}
\item We started the modeling process with a model containing only the
coding time component from telemetry. Explaining 7\% of the variance
we observed in the data, this simple model gave us an
easy-to-understand baseline over which to improve. While we did not
have lines of code measurements available, a slightly more complex
model with it and the developers' seniority at Google explained only
1.7\% of self-reported productivity \cite{murphy}.

\item As a next step, we added all other application telemetry data,
improving the explained variance by a relatively small 2 percentage
points. This hints at the fact that coding time might be the most
relevant predictor among time spent in different applications.

\item From previous studies on productivity we know that rating behavior
differs tremendously between individuals~\cite{meyer,meyer2,ford}.
Consequently, adding a baseline intercept for every developer (their
User ID) made a large difference in the variance the model was able to
explain, to now 34\%. This intercept allows the model to adapt to
different rating behavior among participants---assume one developer's
average rating might be 2.5, while another's might be 4 stars.

\item Adding meeting data from developers' calendars did not further
increase the model's explanatory power. As a result, we left it out in
our best-fit model attempt.

\item For the final model, we added the day attributes developers submitted,
shown in \textbf{\Cref{coeffs}}. The table presents all features with
their coefficients, with the day attributes in the lower half. Many of
the day attributes in the model represent orthogonal pairs, e.g., ``I
slept well'' and ``I did not sleep well.'' Traveling has by far the
largest negative effect on self-reported productivity (down by one
point), followed by being a ``designated response individual,'' having
many interruptions, and a bad sleep. On the plus side, it seems to be
the case that antithetical partners, for example having a good night's
sleep or fewer interruptions than usual, can (almost) entirely make up
for a bad day. ``Designated response individual'' is
``Microspeak''~\cite{microspeak} for ``being on-call.'' It seems
natural that developers lower their reported productivity when
expecting to deal with urgent unforeseen issues. There might be many
reasons for why the attribute ``I worked from home'' (\Cref{mail},
right) did not emerge as significant in the final model, from an
averaging effect in the population (some might prefer it, others not)
to the fact that it might simply make no difference. We originally
included many more team-related attributes (see \Cref{mail}, right)
but they turned out to be insignificant in the final model.
\end{enumerate}

With this final model including day attributes, we were able to
explain almost half (47\%, i.e. $R^2 = 0.47$) of the variance in the observed data. This
final model has a good fit for a concept as complex as
productivity~\cite{munson}. Capturing more would likely be
over-fitting, as the automatic measurements that serve as input to the
model simply do not capture all important aspects of productivity,
such as off-screen work, nor do they give insight into the quality of
the on-screen work, such as the difficulty of a task. With an average
of 18.3 ratings per developer (User ID), over-fitting of the model is
a concern. We thus performed a reduction of our regressed model,
keeping only coefficients that had a statistically significant impact
on the model when left out.

\begin{table}[]
\caption{The features and their coefficient values making up the final productivity
  model.}
\label{coeffs}
  \small
\begin{tabular}{|l|r|}
\hline
\textbf{Feature} & \textbf{Coefficient}\\ \hline
Telemetry Coding time & 0.18\\ \hline
Telemetry Review time & 0.18\\ \hline
Telemetry Other time & 0.09\\ \hline
User ID & {[}-0.93; 1.45{]}\\ \hline
\hline
I was traveling & -0.93\\ \hline
I was a designated response individual & -0.46\\ \hline
I had more interruptions than usual & -0.49\\ \hline
I had fewer interruptions than usual & 0.58\\ \hline
I slept well & 0.31\\ \hline
I did not sleep well & -0.38\\ \hline
I worked longer than usual & 0.24\\ \hline
I worked shorter than usual & -0.35\\ \hline
I had more or longer meetings than usual & -0.26\\ \hline
I had fewer or shorter meetings than usual & 0.23\\ \hline
None of the above applied & -0.32\\ \hline
\end{tabular}
\end{table}

\section{Post-Study Survey}\label{personalized-productivity-reports}

\noindent
One week after the study period, we sent personalized reports to
developers to give them insights into their productivity ratings and
general application use. In the reports, we created a set of
visualizations fusing productivity ratings with automatically measured
product and process. Every developer received a customized report. Our
aim was to test the visualizations for a possible future productivity
dashboard that Microsoft planned to deploy in its Windows development
groups. To assess the effectiveness of the reports and visualizations,
we launched a post-study survey together with the
reports. Participants first saw their personalized report and
afterwards could fill out the short survey. This survey was filled out
by 47 participants (58.0\% of all participants). The sidebar depicts
the layout of the productivity reports and summarizes developers'
opinions of the visualizations.

Besides the visualizations, we were also interested in how the
participants perceived the study in general. Three quarters of
participants would continue to rate their productivity to get an
updated report like the one we send them at the end of the study,
every week.  All of the participants in the survey either liked
(91.2\%) the daily rating emails or were at least neutral (8.2\%)
toward them. Only 4.4\% found the emails at the end of their personal
work day disruptive. The free text participants could enter about the
study anecdotally confirmed this, with participants writing ``I found
it interesting to have this natural point of the day where I would
reflect on how the day went. I started keeping a little work journal
because of it.'' Most developers (71.8\%) agreed or strongly agreed to
the statement that they learned something new from the report. The
majority of developers (89\%) were interested in participating in
future studies on productivity. We take this as a strong sign that
they not only enjoyed participating in the study, but that it provided
tangible benefits to them. Enjoying the study and taking part in the
non-mandatory post-study survey, though, might be correlated---making
these findings susceptible to the volunteer bias.

\begin{table*}[p]\normalsize
\textbf{SIDEBAR: Visualizations in the Productivity Reports}

At the end of the study, each participant received a personal productivity report consisting
mainly of the visualizations A-F below. In a post-study survey, we then assessed the effectiveness
of the visualizations.  The table below depicts an exemplary visualization and the percentage of
respondents who agreed that the respective visualization is easy to understand (Easy), taught them
something new (New), and was actionable (Act). The cells are color-coded on a gradient from pure green,
denoting full agreement, over orange (50\%) to red (0\%), denoting no agreement. As the color
distribution shows, overall, our visualizations tended to be easy to understand, but relatively lower in actionability.
The visualizations of ratings by weekday (B) and application usage by category (D, E) received the highest scores
for easy understanding (85\% and higher); novelty was rated higher for the chart that included the whole study cohort
as a comparison (E) than the non-comparative variant (D). This goes in line with 64\% of study participants agreeing
that ``comparisons with the whole developer population helped me better understand my own values.''
Actionability was below 50\% across the board and the highest for visualizations (D) and (E) at around 45\%. This
highlights both a common problem with visualizations~\cite{vis} and a need to design more actionable visualizations in future reports.

\medskip

\small
\begin{tabular}{|l|p{1.5cm}|p{1.5cm}|p{1.5cm}|}
\hline
\textbf{Visualization Name}                              & \centering\begin{tabular}[c]{@{}l@{}}\textbf{Easy}\end{tabular} & \centering\begin{tabular}[c]{@{}l@{}}\textbf{New}\end{tabular}& \multicolumn{1}{c|}{\textbf{Act}} \\ \hline
\begin{tabular}[c]{@{}p{9cm}@{}}(A) Overall Ratings: distribution of the ratings a
participant submitted\\ \includegraphics[width=5cm]{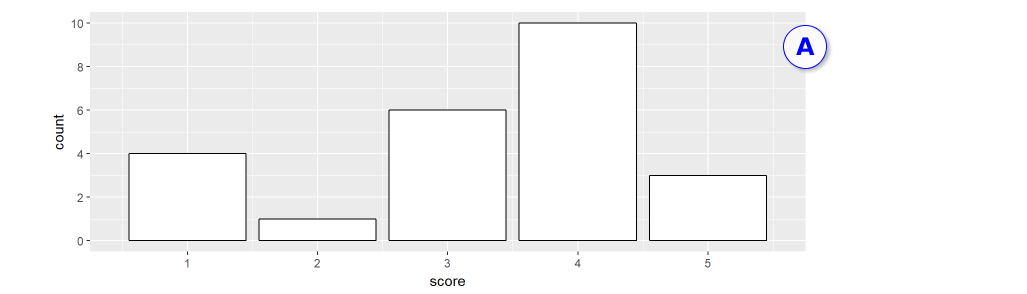}\end{tabular}              & \multicolumn{3}{c|}{\begin{tabular}[c]{@{}l@{}}Not surveyed---purpose was for participants \\ to cross-check data\end{tabular}} \\ \hline

\begin{tabular}[c]{@{}p{9cm}@{}}(B) Ratings per Weekday: average productivity rating per weekday\\ \includegraphics[width=5cm]{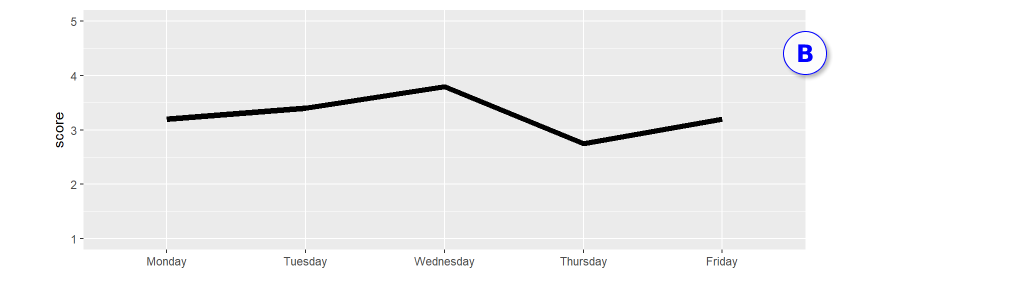}\end{tabular}          & \cellcolor[HTML]{63BE7B}95.4\%                                     & \cellcolor[HTML]{FDCF7E}52.3\%                                       & \cellcolor[HTML]{F8786D}23.2\%                          \\ \hline
\begin{tabular}[c]{@{}p{9cm}@{}}(C) Ratings over Time: daily productivity ratings over time with trend lines\\ \includegraphics[width=5cm]{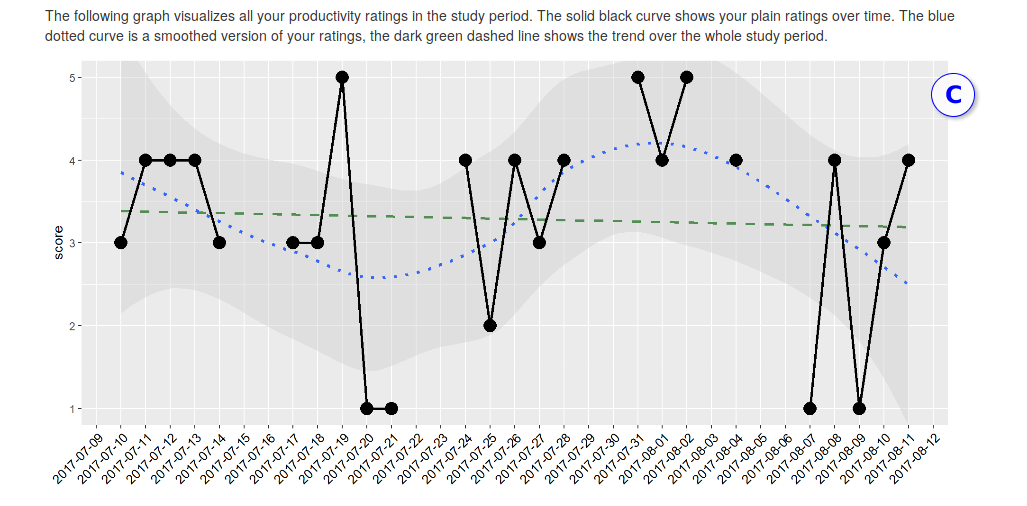}\end{tabular}            & \cellcolor[HTML]{FEE482}59.1\%                                                        & \cellcolor[HTML]{FBAD78}40.9\%                                       & \cellcolor[HTML]{F8696B}18.2\%                          \\ \hline
\begin{tabular}[c]{@{}p{9cm}@{}}(D) Application Usage: polar chart of the developer's average active time across four app categories (dev, browsing, email, other) \\ \includegraphics[width=5cm]{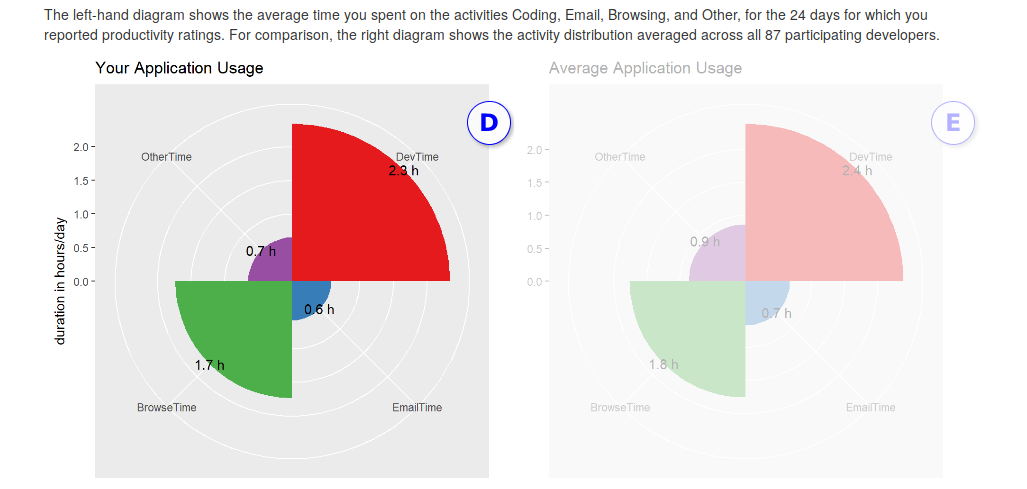}\end{tabular}            & \cellcolor[HTML]{8DCA7E}86.4\%                        & \cellcolor[HTML]{FFEB84}61.4\%                                       & \cellcolor[HTML]{FCC17C}47.7\%                          \\ \hline
\begin{tabular}[c]{@{}p{9cm}@{}}(E) Application Usage Comparison: polar chart of all other developers\\ \includegraphics[width=5cm]{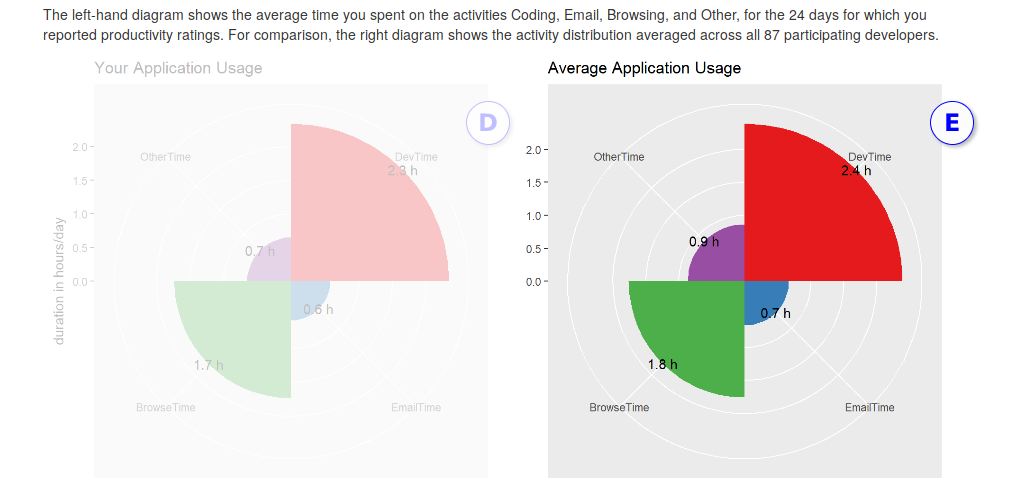}\end{tabular} & \cellcolor[HTML]{8DCB7E}86.3\%                           & \cellcolor[HTML]{C4DA81}74.4\%                                       & \cellcolor[HTML]{FCB77A}44.2\%                          \\ \hline
\begin{tabular}[c]{@{}p{9cm}@{}}(F) Application Usage Over Time: daily active time in app categories\\ \includegraphics[width=5cm]{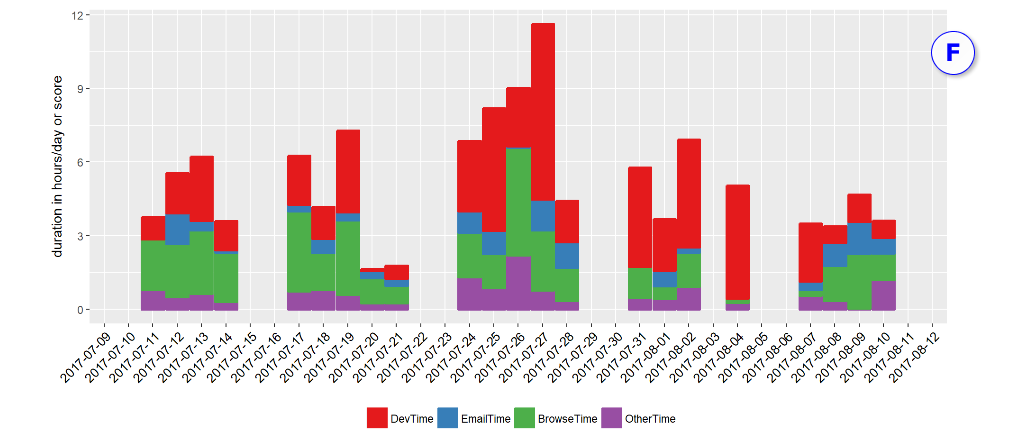}\end{tabular}      & \cellcolor[HTML]{EBE683}65.9\%                                  & \cellcolor[HTML]{FCBA7A}45.4\%                                       & \cellcolor[HTML]{F97D6E}25.0\%                          \\ \hline
\end{tabular}
\end{table*}

\section{Conclusion}

\noindent
This study is an initial step toward bridging the gap between two lines
of thought on productivity: automated measures and self-reported
productivity. With a simple model, we were able to explain almost half of
the variance contained in self-reported productivity when expressed as
automatic product and process measures. The step-wise model fit
confirmed that establishing an individual baseline for rating behavior
is crucial. Time spent coding emerged as an important factor for
developers' self-reported productivity, as did day attributes such as
being able to work with(out) interruptions or sleep quality. While we
need more studies in different contexts to generalize the findings,
our results should at least make organizations aware of the large
conceptual discrepancy between self-reported and measured
productivity.  Another important consideration is that optimizing
for individual productivity is different from optimizing for team
productivity, which might be an ultimately more important metric for
organizations.

A natural extension of our work would be to include measures
such as biometrical data of developers (for example, to quantify
whether they indeed had a good sleep). Since our study ran only for the
course of five weeks and to ease interpretation of the models, we
performed cross-sectional data analysis, disregarding differences in
time. In the future, more complex linear mixed models could enhance the
regression analysis.

Reflecting on our visualizations, we learned that developers preferred
easy-to-grasp visualizations with one clear focus. They seemed to enjoy
a new, unexpected perspective on well-understood data (such as the
weekday chart). Too much detail turned out to be counter-intuitive, even
on the insightfulness of a visualization. By contrast, a comparison with
other developers seems to increase the overall value of a visualization,
but measures need to be taken to preserve the individual's privacy and
not to use this comparative data for performance evaluation, for which
it is not suitable. We initially worried whether our daily emails would
be perceived as disruptive. The opposite was the case: developers
largely embraced them.

Because the study showed that \emph{Coding time} was the most
dominant process measure on self-reported productivity, Microsoft started
investigating to make it a key metric for some of its development teams,
including installing interventions for its developers to increase their
coding time. Finally, most developers learned something new from the
study and almost all would participate in a future study. If by nothing
else, the sheer process of reflecting on their productivity helped
developers be more productive.

\section{ACKNOWLEDGMENT}

\noindent
We thank the Pixou for her Excel magic.
Icons in \Cref{mail} made by Freepik from www.flaticon.com and freepngimg.com.

\section{THREE KEY INSIGHTS}

\begin{itemize}
\item The more time software developers can spend on coding, the higher they rate their productivity.
\item Traveling had the single largest negative impact on self-reported productivity. Quality of sleep and interruption-free work were important, but the weekday was not (no ``Friday effect'').
\item The sheer process of reflecting on their productivity seemingly helped software engineers be more productive.
\end{itemize}






\end{document}